\documentclass[twocolumn]{rps_esrl2020_modified}

\def\papername{\jobname}

\usepackage{amsmath,amsfonts,amssymb,graphicx,bm}
\usepackage{subfigure}
\usepackage{algpseudocode}
\usepackage{algorithm}
\usepackage{multirow}
\usepackage{appendix}

\begin{document}

\markboth{Bai, Huang and Lam}{A DRO Approach to NASA Langley UQ Challenge}

\twocolumn[

\title{A Distributionally Robust Optimization Approach to the NASA Langley Uncertainty Quantification Challenge}

\author{Yuanlu Bai}

\address{Department of Industrial Engineering \& Operations Research, Columbia University, USA. \email{yb2436@columbia.edu}}

\author{Zhiyuan Huang}

\address{Department of Industrial \& Operations Engineering, University of Michigan, Ann Arbor, USA. \email{zhyhuang@umich.edu}}

\author{Henry Lam}

\address{Department of Industrial Engineering \& Operations Research, Columbia University, USA. \email{henry.lam@columbia.edu}}

\begin{abstract} 
	We study a methodology to tackle the NASA Langley Uncertainty Quantification Challenge problem, based on an integration of robust optimization, more specifically a recent line of research known as distributionally robust optimization, and importance sampling in Monte Carlo simulation. The main computation machinery in this integrated methodology boils down to solving sampled linear programs. We will illustrate both our numerical performances and theoretical statistical guarantees obtained via connections to nonparametric hypothesis testing. 	
\end{abstract}


]


We consider the NASA Langley Uncertainty Quantification (UQ) Challenge problem (\cite{nasa}) where, given a set of ``output" data and under both aleatory and epistemic uncertainties, we aim to infer a region that contains the true values of the associated variables. These steps allow us to investigate the reduction of uncertainty by obtaining further information and estimate the failure probabilities of related systems. To tackle these challenges, we study a methodology based on an integration of robust optimization (RO), more specifically, a recent line of research known as \emph{distributionally robust optimization (DRO)}, and importance sampling in Monte Carlo simulation. We will see that the main computation machinery in this integrated methodology boils down to solving sampled linear programs (LPs). In this paper, we will explain our methodology, introduce some theoretical statistical guarantees via connections to nonparametric hypothesis testing, and summarize the numerical results on the UQ Challenge. 


\section{Overview of Our Methodology (Problem A)}
We first give a high-level overview of our methodology in extracting a region $E$ that contains the true epistemic variables. For convenience, we call this region an ``eligibility set" of $e$. For each value of $e$ inside $E$, we also have a set (in the space of probability distributions) that contains ``eligible" distributions for the random variable $a$. For the sake of computational tractability (as we will see shortly), the eligibility set of $e$ is represented by a set of sampled points in $E_0$ that approximate its shape, whereas the eligibility set of $a$ is represented by probability weights on sampled points on $A$. The eligibility set $E$ and the corresponding eligibility set of distributions for $a$ are obtained by solving an array of LPs that are constructed from the properly sampled points, and then deciding eligibility by checking the LP optimal values against a threshold that resembles the ``$p$-value" approach in hypothesis testing. This methodology involves a dimension-collapsing transformation $\mathbf S$, applied on the raw data, which ultimately allows using the Kolgomorov-Smirnov (KS) statistic to endow rigorous statistical guarantees. Algorithm \ref{algo1} is a procedural description of our approach to construct the eligibility set $E$, and it also gives as a side product an eligibility set of the distributions of $a$ for each $e$, represented by weights in the set \eqref{eligibility a}. In the following, we explain the elements and terminologies in this algorithm in detail. 

\setlength{\textfloatsep}{0.1cm}
\begin{algorithm}[b]
	\caption{Constructing eligibility set $E$}
	\textbf{Input: }Data $D_1=\{(y^{(i)}(t))_{t=0,\ldots,T}\}_{i=1,\ldots,n_1}$. A uniformly sampled set of $e^{(l)},l=1,\ldots,n_2$ over $E_0$. A uniformly sampled set of $a^{(r)},r=1,\ldots,k$ over $A$. A summary function $\mathbf S(\cdot):\mathbb R^{n_t+1}\to\mathbb R^m$. A target confidence level $1-\alpha$.
	
	\textbf{Procedure: }
	\begin{algorithmic}
		
		\State \textbf{1. Simulate outputs from the baseline distribution:} Evaluate $(y(a^{(r)},e^{(l)},t))_{t=0,\ldots,T}$ for $r=1,\ldots,k$, $l=1,\ldots,n_2$.

		\State \textbf{2. Summarize the outputs:} Evaluate $\mathbf s^{(i)}=\mathbf S((y^{(i)}(t))_{t=0,\ldots,T})$ for $i=1,\ldots,n_1$, and $\mathbf S(y(a^{(r)},e^{(l)},t))_{t=0,\ldots,T})$ for $r=1,\ldots,k$, $l=1,\ldots,n_2$.
		
		\State \textbf{3. Compute the degree of eligibility:} For each $l=1,\ldots,n_2$, solve optimization problem Eq. \eqref{alternate} to obtain $q_l^*$. 
		
		\State \textbf{4. Construct the eligibility set:} Output $E=\{e^{(l)}:q_l^*\leq q_{1-\alpha/m}\}$. Smooth the set if needed.
		
	\end{algorithmic}\label{algo1}
\end{algorithm}

\section{A DRO Perspective}\label{sec:DRO}
Our starting idea is to approximate the set
\begin{equation}
E=\{e\in E_1:\text{there exists\ }P_e\text{\ s.t.\ }d(P_e,\hat P)\leq\eta\}\label{E}
\end{equation}
where $P_e$ is the probability distribution of $\{y(a,e,t)\}_{t=0,\ldots,T}$, namely the outputs of the simulation model $\{y(a,e,t)\}_{t=0,\ldots,T}$ at a fixed $e$ but random $a$. $\hat P$ denotes the empirical distribution of $D_1$, more concretely the distribution given by
$$\hat P(\cdot)=\frac{1}{n_1}\sum_{i=1}^{n_1}\delta_{(y^{(i)}(t))_{t=0,\ldots,T}}(\cdot)$$
where $\delta_{(y^{(i)}(t))_{t=0,\ldots,T}}(\cdot)$ denotes the Dirac measure at $(y^{(i)}(t))_{t=0,\ldots,T}$. $d(\cdot,\cdot)$ denotes a discrepancy between two probability distributions, and $\eta\in\mathbb R_+$ is a suitable constant. Intuitively, $E$ in Eq. \eqref{E} is the set of $e$ such that there exists a distribution for the outputs that is close enough to the empirical distribution from the data. If for a given $e$ there does not exist any possible output distribution that is close to $\hat P$, then $e$ is likely not the truth. The following gives a theoretical justification for using Eq. \eqref{E}:
\begin{theorem}
	Suppose that the true distribution of the output $(y(t))_{t=0,\ldots,T}$, called $P_{true}$, satisfies $d(P_{true},\hat P)\leq\eta$ with confidence level $1-\alpha$, i.e., we have 
	\begin{equation}
	\mathbb P(d(P_{true},\hat P)\leq\eta)\geq1-\alpha\label{elementary}
	\end{equation}
	where $\mathbb P$ denotes the probability with respect to the data. Then the set $E$ in Eq. \eqref{E} satisfies $\mathbb P(e_{true}\in E)\geq1-\alpha$, where $e_{true}$ denotes the true value of $e$. Similar deduction holds if Eq. \eqref{elementary} holds asymptotically (as the data size grows), in which case the same asymptotic modification holds for the conclusion. \label{basic guarantee}
\end{theorem}
The proof of Theorem \ref{basic guarantee} is straightforward. 
	Note that $d(P_{true},\hat P)\leq\eta$ implies $e_{true}\in E$. Thus we have $\mathbb P(e_{true}\in E)\geq\mathbb P(d(P_{true},\hat P)\leq\eta)\geq1-\alpha$. Similar derivation holds for the asymptotic version. 
    
    In Eq.~\eqref{E}, the set of distributions $\{P_e:d(P_e,\hat P)\leq\eta\}$ is analogous to the so-called uncertainty set or ambiguity set in the RO literature (e.g., \cite{bertsimas2011theory,ben2002robust}), which is a set postulated to contain the true values of uncertain parameters in a model. RO generally advocates decision-making under uncertainty that hedges against the worst-case scenario, where the worst case is over the uncertainty set (and thus often leads to a minimax optimization problem). DRO, in particular, focuses on problems where the uncertainty is on the probability distribution of an underlying random variable (e.g., \cite{wiesemann2014distributionally,delage2010distributionally}). This is the perspective that we are taking here, where $a$ has a distribution that is unknown, in addition to the uncertainty on $e$. Moreover, we also take a generalized view of RO or DRO here as attempting to construct an eligibility set of $e$ instead of finding a robust decision via a minimax optimization. 
    
    Theorem \ref{basic guarantee} focuses on the situation where the uncertainty set is constructed and calibrated from data, which is known as data-driven RO or DRO (\cite{bertsimas2018data,hong2017learning}). If such an uncertainty set has the property of being a confidence region for the uncertain parameters or distributions, then by solving RO or DRO, the confidence guarantee can be translated to the resulting decision, or the eligibility set in our case. Here we have taken a \emph{nonparametric} and \emph{frequentist} approach, as opposed to other potential Bayesian methods.
    

In implementation we choose $\alpha=0.05$, so that the eligibility set $E$ has the interpretation of approximating a $95\%$ confidence set for $e$. In the above developments, $d(P_e,\hat P)\leq\eta$ can in fact be replaced with more general set $P_e\in\mathcal U$ where $\mathcal U$ is calibrated from the data. Nonetheless, the distance-based set (or ``ball") surrounding the empirical distribution is intuitive to understand, and our specific choice of the set below falls into such a representation.

To use Eq.~\eqref{E}, there are two immediate questions:
\begin{enumerate}
	\item What $d(\cdot,\cdot)$ should and can we use, and how do we calibrate $\eta$?
	\item How do we determine whether there exists $P_e$ that satisfies $d(P_e,\hat P)\leq\eta$ for a given $e$?
\end{enumerate}

For the first question, we first point out that in theory many choices of $d$ could be used (basically, any $d$ that satisfies the confidence property in Theorem \ref{basic guarantee}). But, a poor choice of $d$ would lead to a more conservative result, i.e., larger $E$, than others. A natural choice of $d$ should capture the discrepancy of the distributions efficiently. Moreover, the choice of $d$ should also account for the difficulty in calibrating $\eta$ such that the assumption in Theorem \ref{basic guarantee} can be satisfied, as well as the computational tractability in solving the eligibility determination problem in Eq.~\eqref{E}.

Based on the above considerations, we construct $d$ and calibrate $\eta$ as follows. First, we ``summarize" the data $D_1$ into a lower-dimensional representation, say $\{s_1^{(i)},\ldots,s_m^{(i)}\},i=1,\ldots,n_1$, where $s_r^{(i)}=S_r({y^{(i)}(t)}_{t=0,\ldots,T})$ for some function $S_r(\cdot)$. For convenience, we denote $\mathbf S(\cdot)=(S_1(\cdot),\ldots,S_m(\cdot)):\mathbb R^{n_t+1}\to\mathbb R^m$, and $\mathbf s^{(i)}=(s_1^{(i)},\ldots,s_m^{(i)})$. We call $\mathbf S(\cdot)$ the ``summary function" and $\mathbf s^{(i)}$ the ``summaries" of the $i$-th output. $\mathbf S(\cdot)$ attempts to capture important characteristics of the raw data (we will see later that we use the positions and values of the peaks extracted from Fourier analysis). Also, the low dimensionality of $\mathbf s^{(i)}$ is important to calibrate $\eta$ well.



Next, we define
\begin{equation}
d(P_e,\hat P)=
\max_{r=1,\ldots,m}\sup_{x\in\mathbb R}\left|F_{e,r}(x)-\hat F_r(x)\right|\label{KS}
\end{equation}
where $\hat F_r(x)=\frac{1}{n_1}\sum_{i=1}^{n_1}I(x\leq s_r^{(i)})$, with $I(\cdot)$ denoting the indicator function, is the empirical distribution function of $s_r^{(i)}$ (i.e., the distribution function of $\hat P$ projected onto the $r$-th summary). $F_{e,r}(x)$ is the probability distribution function of the $r$-th summary of the simulation model output $S_r(y(a,e,t))_{t=0,\ldots,T}$ (i.e., the distribution function of the projection of $P_e$ onto the $r$-th summary). We then choose $\eta=q_{1-\alpha/m}/\sqrt{n_1}$ as the $(1-\alpha/m)$-quantile of the Kolmogorov-Smirnov (KS) statistic, namely that $q_{1-\alpha/m}$ is the $(1-\alpha/m)$-quantile of $\sup_{x\in[0,1]}BB(x)$ where $BB(\cdot)$ denotes a standard Brownian bridge. 

To understand Eq.~\eqref{KS}, note that the set of $P_e$ that satisfies $d(P_e,\hat P)\leq\eta$ is equivalent to $P_e$ that satisfies
\begin{equation}
\sup_{x\in\mathbb R}\left|F_{e,r}(x)-\hat F_r(x)\right|\leq\frac{q_{1-\alpha/m}}{\sqrt{n_1}},\ \ r=1,\ldots,m\label{elaborate constraint}
\end{equation}Here, $\sup_{x\in\mathbb R}\left|F_{e,r}(x)-\hat F_r(x)\right|$ is the KS-statistic for a goodness-of-fit test against the distribution $F_{e,r}(x)$, using the data on the $r$-th summary. Since we have $r$ summaries and hence $r$ tests, we use a Bonferroni correction and deduce that
 $$
\begin{aligned}
\liminf_{n_1\to\infty}
\mathbb P\bigg(&\sup_{x\in\mathbb R}\left|F_{true,r}(x)-\hat F_r(x)\right|\leq\frac{q_{1-\alpha/m}}{\sqrt{n_1}},\\
&\ r=1,\ldots,m\bigg)
\geq1-\alpha\
\end{aligned}
$$
where $F_{true,r}$ denotes the true distribution function of the $r$-th summary. Thus, the (asymptotic version of the) assumption in Theorem \ref{basic guarantee} holds. Note that here the quality of the summaries does not affect the statistical correctness of our method (in terms of overfitting), but it does affect crucially the resulting conservativeness (in the sense of getting a larger $E$). Moreover, in choosing the number of summaries $m$, there is a tradeoff between the conservativeness coming from \emph{representativeness} and \emph{simultaneous estimation}. On one end, using more summaries means more knowledge we impose on $P_e$, which translates into a smaller feasible set for $P_e$ and ultimately a smaller eligible set $E$. This relation, however, is true only if there is no statistical noise coming from the data. In the case of finite data size $n_1$, then more summaries also means that constructing the feasible set for $P_e$ requires more simultaneous estimations in calibrating its size, which is manifested in the Bonferroni correction whose degree increments with each additional summary. In our implementation (see Section \ref{sec:summary Fourier}), we find that using 12 summaries seems to balance well this representativeness versus simultaneous estimation error tradeoff.


Now we address the second question on how we can decide, for a given $e$, whether a $P_e$ exists such that $d(P_e,\hat P)\leq\eta$.
We first rephrase the representation 
with a change of measure. Consider a ``baseline" probability distribution, say $P_0$, that is chosen by us in advance. A reasonable choice, for instance, is the uniform distribution over $A$, the support of $a$. Then we can write $d(P_e,\hat P)\leq\eta$
as
\begin{equation}
\sup_{x\in\mathbb R}\left|\int_{S_r(u)\leq x}W_e(u)dP_0(u)-\hat F_r(x)\right|\leq\frac{q_{1-\alpha/m}}{\sqrt{n_1}}
\label{elaborate constraint1}
\end{equation}
for $r=1,\ldots,m$ where $W_e(\cdot)=dP_e/dP_0$ is the Radon-Nikodym derivative of $P_e$ with respect to $P_0$, and we have used the change-of-measure representation $F_{e,r}(x)=\int_{S_r(u)\leq x}W_e(u)dP_0(u)$. Here we have assumed that $P_0$ is suitably chosen such that absolute continuity of $P_e$ with respect to $P_0$ holds. Eq.~\eqref{elaborate constraint1} turns the determination of the existence of eligible $P_e$ into the existence of an eligible Radon-Nikodym derivative $W_e(\cdot)$. 



The next step is to utilize Monte Carlo simulation to approximate $P_0$. More specifically, given $e$, we run $k$ simulation runs under $P_0$ to generate $(y(a^{(j)},e,t))_{t=0,\ldots,T}$ for $j=1,\ldots,k$. Then Eq.~\eqref{elaborate constraint1} can be approximated by
\begin{equation}
\begin{aligned}
\sup_{x\in\mathbb R}\Bigg|&\sum_{j=1}^kW_jI(S_r((y(a^{(j)},e,t))_{t=0,\ldots,T})\leq x)-\\
&\hat F_r(x)\Bigg|\leq\frac{q_{1-\alpha/m}}{\sqrt{n_1}},\ r=1,\ldots,m\label{elaborate constraint2}
\end{aligned}
\end{equation}
where $W_j=(1/k)(dP_e/dP_0((y(a^{(j)},e,t))))$ represents the (unknown) sampled likelihood ratio from the view of importance sampling (\cite{blanchet2012state}; \cite{asmussen2007stochastic} Chapter 5). Our task is to find a set of weights, $W_j,j=1,\ldots,k$, such that Eq.~\eqref{elaborate constraint2} holds. These weights should approximately satisfy the properties of the Radon-Nikodym derivative, namely positivity and integrating to one. Thus, we seek for $W_j,j=1,\ldots,k$ such that
\begin{eqnarray}
&&
\begin{aligned}
\sup_{x\in\mathbb R}\Bigg|&\sum_{j=1}^kW_jI(S_r((y(a^{(j)},e,t))_{t=0,\ldots,T})\leq x)-\\
&\hat F_r(x)\Bigg|\leq\frac{q_{1-\alpha/m}}{\sqrt{n_1}},\ r=1,\ldots,m
\end{aligned}\label{elaborate constraint KS}\\
&&\sum_{j=1}^kW_j=1,\ W_j\geq0\ \text{\ for\ }j=1,\ldots,k\label{elaborate constraint3}
\end{eqnarray}
where Eq.~\eqref{elaborate constraint3} enforces the weights to lie in a probability simplex. If 
$k$ is much larger than 
$n_1$, then the existence of $W_j,j=1,\ldots,k$ satisfying Eq.~\eqref{elaborate constraint KS} and Eq.~\eqref{elaborate constraint3} would determine that the considered $e$ is in $E$. 
To summarize, we have:
\begin{theorem}
	Suppose $k=\omega(n_1)$, and $P_{true}$ is absolutely continuous with respect to $P_0$ and that $\|dP_{true}/dP_0\|_\infty\leq C$ for some constant $C>0$ and $\|\cdot\|_\infty$ denotes the essential supremum. Suppose, for each $e$, we generate $k$ simulation replications to get $(y(a^{(j)},e,t))_{t=0,\ldots,T}),j=1,\ldots,k$, where $a^{(j)}$ are drawn from $P_0$ in an i.i.d. fashion. Then the set
    \begin{eqnarray*}
        E&=&\Big\{e:\text{there exists }W_j,j=1,\ldots,k\text{ such that}\\
        &&\text{Eq.~\eqref{elaborate constraint KS} and Eq.~\eqref{elaborate constraint3} hold}\Big\}
    \end{eqnarray*}
	will satisfy
	$$\liminf_{n_1\to\infty,k/n_1\to\infty}\mathbb P(e_{true}\in E)\geq1-\alpha$$\label{main guarantee}
\end{theorem}
Note that in Theorem \ref{main guarantee}, $W_j$'s represent the unknown sampled likelihood ratios such that, together with the $a^{(j)}$'s generated from $P_0$, the function$\sum_{j=1}^kW_jI(S_r((y(a^{(j)},e,t))_{t=0,\ldots,T})\leq\cdot)$ approximates the unknown true $r$-th summary distribution function $F_{true,r}$. To use the above $E$ and elicit the guarantee in Theorem \ref{main guarantee}, we still need some steps in order to conduct feasible numerical implementation. First, we need to discretize or sufficiently sample $e$'s over $E_0$, since checking the existence of eligible $W_j$'s for all $e$ is computationally infeasible. In our implementation we draw $n_2=1000$ $e$'s uniformly over $E_0$, call them $e^{(1)},\ldots,e^{(n_2)}$, and then put together the geometry of $E$ from the eligible $e^{(l)}$'s. Second, the current representation of the KS constraint Eq.~\eqref{elaborate constraint KS} 
involves entire distribution functions. 
We can write Eq.~\eqref{elaborate constraint KS} as a finite number of linear constraints, given by
\begin{equation}
\begin{aligned}
&\hat F_r(s_r^{(i)}+)-\frac{q_{1-\alpha/m}}{\sqrt{n_1}}\\
\leq&\sum_{j=1}^kW_jI(S_r((y(a^{(j)},e,t))_{t=0,\ldots,T})
\leq s_r^{(i)})\\
\leq&\hat F_r(s_r^{(i)}-)+\frac{q_{1-\alpha/m}}{\sqrt{n_1}}
\end{aligned}\label{feasibility partial}
\end{equation}
for $i=1,\ldots,n_1, r=1,\ldots,m$ where $s_r^{(i)},i=1,\ldots,n_1$ are the $r$-th summary of the $i$-th data point, and $s_r^{(i)}+$ and $s_r^{(i)}-$ denote the right and left limits of the empirical distribution at 
$s_r^{(i)}$.

Thus, putting everything together, we solve, for each $e^{(l)},l=1,\ldots,n_2$, the feasibility problem: \emph{Find $W_j,j=1,\ldots,k$ such that Eq.~\eqref{feasibility partial} and Eq.~\eqref{elaborate constraint3} hold}.
If there exists feasible $W_j,j=1,\ldots,k$, then $e^{(l)}$ is eligible. The set $\{e^{(l)}:e^{(l)} \text{\ is eligible}\}$ is an approximation of $E$. Note that this is a ``sampled" subset of $E$. In general, without running the simulation at the other points of $E$, there is no guarantee whether these other points are eligible or not. However, if the distribution of $\{y(a,e,t)\}_{t=0,\ldots,T}$ is continuous in $e$ in some suitable sense, then it is reasonable to believe that the neighborhood of an eligible point $e^{(l)}$ is also eligible (and vice versa). In this case, we can ``smooth" the discrete set of $\{e^{(l)}:e^{(l)} \text{\ is eligible\ }\}$ if needed (e.g., by doing some clustering and taking the convex hull of each cluster). Note that the feasibility problem above is a linear problem in the decision variables $W_j$'s.

Lastly, we offer an equivalent approach to the above procedure that allows further flexibility in choosing the threshold $q_{1-\alpha/m}$, which currently is set as the Bonferroni-adjusted KS critical value. This equivalent approach leaves this choice of threshold open and can determine the set of eligible $e^{(l)}$ as a function of the threshold, thus giving some room to improve conservativeness should the formed approximate $E$ turns out to be too loose according to other expert opinion. Here, we solve, for each $e^{(l)},l=1,\ldots,n_2$, the optimization problem
\begin{align}
\begin{array}{lll}
q_l^*=&\min&q\\
&\text{s.t.}&
\hat F_r(s_r^{(i)}+)-\frac{q}{\sqrt{n_1}}\\
&&\leq\sum_{j=1}^kW_j\times\\
&&I(S_r((y(a^{(j)},e^{(l)},t))_{t=0,\ldots,T})\leq s_r^{(i)})\\
&&\leq\hat F_r(s_r^{(i)}-)+\frac{q}{\sqrt{n_1}}\\
&&\text{for } i=1,\ldots,n_1, r=1,\ldots,m;\\
&&\sum_{j=1}^kW_j=1,\ W_j\geq0\\
&&\text{for\ }j=1,\ldots,k
\end{array}\label{alternate}
\end{align}
where the decision variables are $W_j,j=1,\ldots,k$ and $q$. If the optimal value $q_l^*$ satisfies $q_l^*\leq q_{1-\alpha/m}$, then $e^{(l)}$ is eligible (This can be seen by checking its equivalence to the feasibility problem via the monotonicity of the feasible region for $W_j$'s in Eq.~\eqref{alternate} as $q$ increases). The rest then follows as above that $\{e^{(l)}:e^{(l)} \text{\ is eligible}\}$ is an approximation of $E$. Like before, Eq.~\eqref{alternate} is an LP. Moreover, here $q^*_l$ captures in a sense the ``degree of eligibility" of $e^{(l)}$, and allows convenient visualization by plotting $q_l^*$ against $e^{(l)}$ to assess the geometry of $E$. For these reasons we prefer to use Eq.~\eqref{alternate} over the feasibility problem before. These give the full procedure in Algorithm \ref{algo1}.

Finally, we also present how to find eligible distributions of $a$ for an eligible $e^{(l)}$. The set of eligible distributions of $a$ is approximated by the weights $W_j$'s that satisfy Eq.~\eqref{feasibility partial} and Eq.~\eqref{elaborate constraint3}, namely
\begin{align}
&\Bigg\{W_j,j=1,\ldots,k:\hat F_r(s_r^{(i)}+)-\frac{q_{1-\alpha/m}}{\sqrt{n_1}}\notag\\
&\leq\sum_{j=1}^kW_jI(S_r((y(a^{(j)},e^{(l)},t))_{t=0,\ldots,T})
\leq s_r^{(i)})\notag\\
&\leq\hat F_r(s_r^{(i)}-)+\frac{q_{1-\alpha/m}}{\sqrt{n_1}},\ i=1,\ldots,n_1, r=1,\ldots,m\notag\\
&\sum_{j=1}^kW_j=1,\ W_j\geq0\ \text{\ for\ }j=1,\ldots,k\Bigg\}\label{eligibility a}
\end{align}
where $W_j$ is the probability weight on $a^{(j)}$. From this, one could also obtain approximate bounds for quantities related to the distribution of $a$. For instance, to get approximate bounds for the mean of $a$, we can maximize and minimize $\sum_jW_ja^{(j)}$ subject to constraint \eqref{eligibility a}. 


To close this section, we discuss some related literature to our methodology that is not yet mentioned. In our development, we have constructed an uncertainty set for the unknown distribution $P_e$ via a confidence region associated with the KS goodness-of-fit test. This uncertainty set has been proposed in \cite{bertsimas2018robust}, and other distance-based uncertainty sets, including $\phi$-divergence (\cite{ben2013robust}) and Wasserstein distance (\cite{esfahani2018data}), have also been used. We use a simultaneous group of KS statistics with Bonferroni correction, motivated by the tractability in the resulting integration with the importance weighting. The closest work to our framework is the stochastic simulation inverse calibration problem studied in \cite{goeva2019optimization}, but they consider single-dimensional output and parameter to calibrate the input distributions, in contrast to our ``summary" approach via Fourier analysis and the multi-dimensional settings we face. Finally, we point out that the use of simulation and importance sampling in robust optimization has also been studied in risk quantification in operations research and mathematical finance (e.g., \cite{glasserman2014robust,ghosh2019robust,lam2013robust}).

In the remainder of this paper, we will illustrate the use of our methodology and report briefly our numerical results for the UQ Challenge.

\section{Summarizing Discrete-Time Histories using Fourier Transform}\label{sec:summary Fourier}
By observing the plot of the outputs $y^{(i)},i=1,\dots,n_1$, we judge that these time series are highly seasonal. Naturally, we choose to use Fourier transform to summarize $(y(t))_{t=0,\dots,T}$, and we may write $y(t)$ in the form $
y(t)=\sum_{k=-\infty}^{\infty}C_k e^{-ik\omega_0 t}.
$
\par
\begin{figure}[h]
\vspace{-5mm}
    \centering
    \subfigure[Real part]{
    \includegraphics[width=0.22\textwidth]{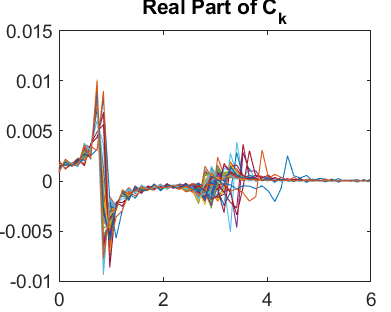}
    }
    \subfigure[Imaginary part]{
    \includegraphics[width=0.22\textwidth]{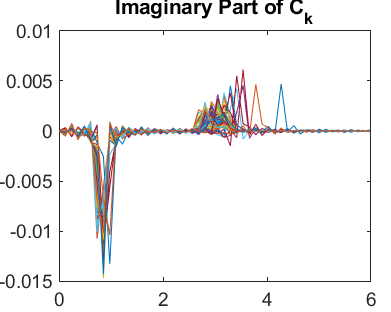}
    }
    \caption{The real part and the imaginary part pf $C_k$'s against the corresponding frequencies}
    \label{real and imag}
    \vspace{-5mm}
\end{figure}

 
First we try to apply Fourier transform to $y^{(i)},i=1,\dots,n_1$. For each $y^{(i)}$, we compute the $C_k$'s. Fig. \ref{real and imag} shows the real part and the imaginary part of $C_k$'s against the corresponding frequencies. 

For the real part, we see that there is a large positive peak, a large negative peak, a small positive peak and a small negative peak. After testing, we confirm that for any $i$, the large peaks lie in the first 14 terms (from 0Hz to 1.59Hz), while the small peaks lie between the 15th term and the 50th term (from 1.71Hz to 5.98Hz). For the imaginary part, we see that there is a large negative peak and a small positive peak. The large peak is also located in the first 14 terms and the small peak between the 15th term and the 50th one. 
\par 
Therefore, we choose to use the following method to summarize $y$ (i.e., construct the function $\textbf{S}(\cdot)$): first, we apply the Fourier transform to compute $C_k$'s and the corresponding frequencies; second, we compute the real part and the imaginary part of $C_k$'s; third, for the real part, we find the maximum value and the minimum value over $[0Hz, 1.59Hz]$ and $[1.71Hz, 5.98Hz]$, as well as their corresponding frequencies; fourth, for the imaginary part, we find the minimum value over $[0Hz, 1.59Hz]$ and the maximum value over $[1.71Hz, 5.98Hz]$ as well as their corresponding frequencies. Then we use these 12 parameters as the summaries of $y$. 
\par
\begin{figure}[h]
\vspace{-5mm}
	\centering
	\includegraphics[width=5cm]{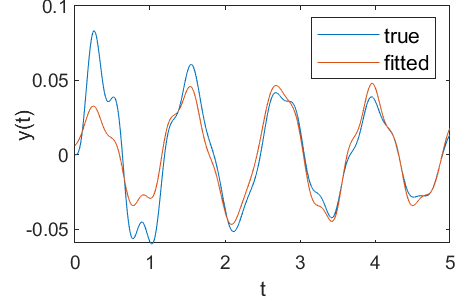}\\
	\caption{Fitting $y^{(1)}$ with the 12 parameters}
	\label{fitting}
	\vspace{-8mm}
\end{figure}

 
To illustrate how well these summaries fit $y$, Fig. \ref{fitting} shows the comparison for $y^{(1)}$. The fit qualities of other time series are similar to this example. Though they may not be extremely close to each other, the fitted curves do resemble the original curves. Note that it is entirely possible to improve the fitting if we keep more frequencies even if they are not as significant as the main peaks. 
On the other hand, as discussed in Section \ref{sec:DRO}, using a larger number of summaries both represents more knowledge of $P_e$ (better fitting) but also leads to more simultaneous estimation error when using the Bonferroni correction needed in calibrating the set for $P_e$. To balance the conservativeness of our approach coming from representativeness versus simultaneous estimation, we choose to use the 12-parameter summaries depicted before. 

\section{Uncertainty Reduction (Problem B)}
\subsection{Ranking Epistemic Parameters (B.1 and B.2)}\label{sec:UR}
Now we implement Algorithm \ref{algo1} with $n_2=k=1000$ and the summary function $\mathbf{S}(\cdot)$ defined in the previous section. The dimension of the summary function is $m=12$. We choose $\alpha$ to be 0.05. Thus, following the algorithm, for each $l=1,\dots,n_2$, we compute $q_l^*$ and then compare it with $q_{1-\alpha/m}=q_{1-0.05/12}=1.76$. 
\par
\begin{figure}[h]
    \vspace{-8mm}
    \centering
    \subfigure[$e_1$]{
    \includegraphics[width=0.22\textwidth]{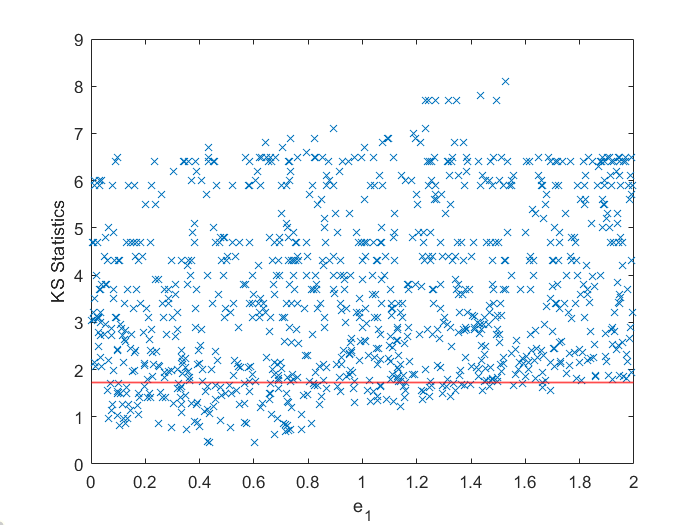}
    }
    \subfigure[$e_2$]{
    \includegraphics[width=0.22\textwidth]{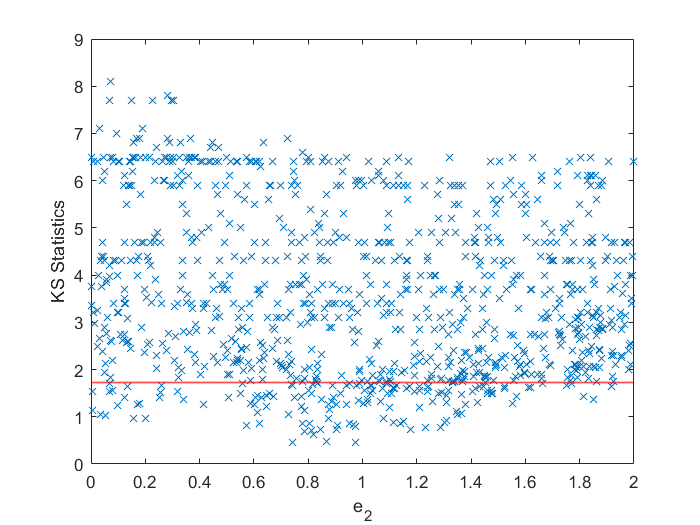}
    }
    \subfigure[$e_3$]{
    \includegraphics[width=0.22\textwidth]{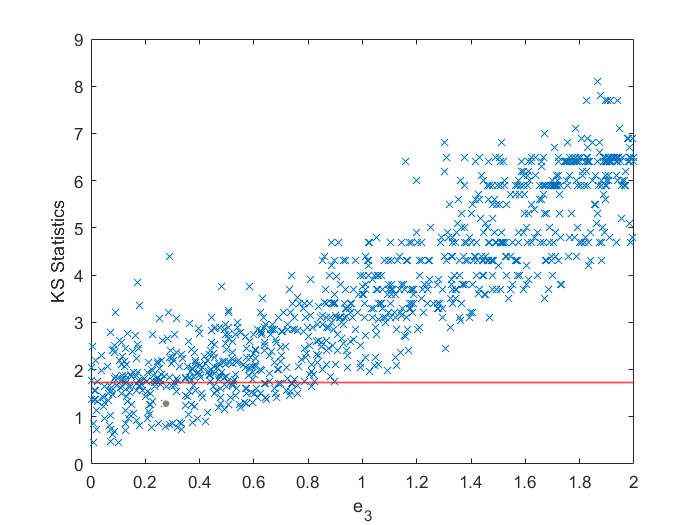}
    }
    \subfigure[$e_4$]{
    \includegraphics[width=0.22\textwidth]{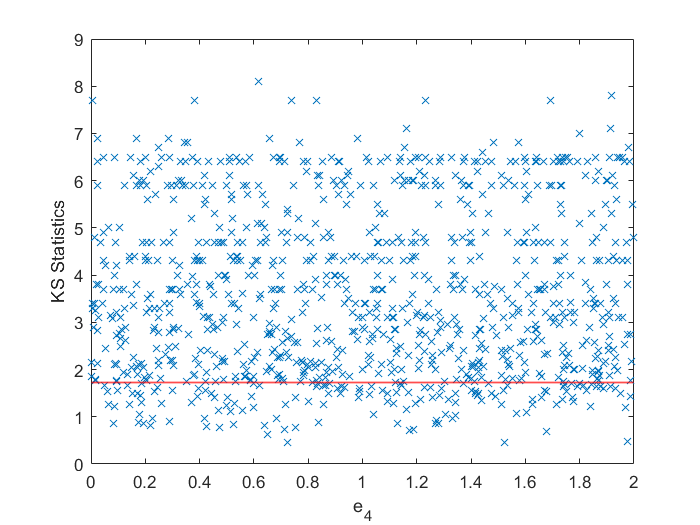}
    }
    \caption{$q_l^*$ against each epistemic variable}
    \label{e}
    \vspace{-8mm}
\end{figure}

 
In Fig. \ref{e}, we plot the $q_l^*$'s against each dimension of $e$. The red horizontal lines in the graphs correspond to $q_{1-\alpha/m}=1.76$. Thus the dots \emph{below} the red lines constitute the eligible $e$'s. We rank the epistemic parameters according to these graphs, namely we rank higher the parameter whose range can potentially be reduced the most. Note that this ranking scheme can be summarized using more rigorous metrics related to the expected amount of eligible $e$'s after range shrinkage, but since there are only four dimensions, using the graphs directly seem sufficient for our purpose here.
\par 
We find that the values of $e_2$ and $e_4$ of the eligible $e$'s broadly range from 0 to 2, which implies that reducing the ranges of these two dimensions could hardly reduce our uncertainty. By contrast, the values of $e_1$ and $e_3$ of the eligible $e$'s are both concentrated in the lower part of $[0,2]$. Thus, our ranking of the epistemic parameters according to their ability to improve the predictive ability is $e_3>e_1>e_2>e_4$.
\par 
Chances are that the true values of $e_1$ and $e_3$ are relatively small. In order to further pinpoint the true values of $e_1$ and $e_3$, we choose to make two uncertainty reductions: increase the lower limits of the bounding interval of $e_1$ and $e_3$.

\subsection{Impact of the value of $n_1$ (A.2)}
To investigate the impact of the value of $n_1$, 
for different values of $n_1$ we randomly sample $n_1$ outputs without replacement. Then we take these outputs as the new data set. By repeating implementing Alg. \ref{algo1}, we find that the larger is $n_1$, the smaller is the proportion of eligible $e$'s. It is intuitive that as the data size grows, $e$ can be better pinpointed. Moreover, except for $e_4$, the range of each epistemic variable of eligible $e$'s obviously shrinks as $n_1$ increases, which further confirms that $e_4$ is the least important epistemic variable. 
\subsection{Updated Parameter Ranking (B.3)}
After the epistemic space is reduced, we repeat the process in Section \ref{sec:UR} but now $e$'s are generated uniformly from $E_1$. From the associated scatter plots (not shown here due to space limit), the updated ranking of the epistemic parameters is $e_2>e_3>e_1>e_4$.


\section{Reliability of Baseline Design (Problem C)}
\subsection{Failure Probabilities and Severity (C.1, C.2 and C.5)}\label{sec:c1c2}
Combining the refined range of $e$ provided by the host with our Algorithm \ref{algo1}, we construct $E\subset E_1$. To estimate 
$\min_{e\in E}/\max_{e\in E}\mathbb{P}(g_i(a,e,\theta)\geq 0)$, we run simulations to respectively solve
\begin{equation}
    \begin{split}
    \min/\max\ & \sum_{j=1}^k W_j I(g_i(a^{(j)},e,\theta)\geq 0)\\
    \text{s.t. } & e\in E, W\in U
    \end{split}
\end{equation}
where $U$ is the set of $(W_1,\cdots,W_k)$ in Eq.~\eqref{eligibility a}. These give the range of $R_i(\theta)$. We use the same method to approximate $R(\theta)$, the failure probability for any requirement. Note that in our implementation the $E$ in the formulations above is represented by discrete points $e^{(l)}$'s. As discussed previously, under additional smoothness assumptions, we could ``smooth" these points to obtain a continuum. Nonetheless, under sufficient sampling of $e^{(l)}$, the discretized set should be a good enough approximation in the sense that the optimal values from the ``discretized" problems are close to those using the continuum. 
\par 
Using the above method, we get that the ranges of $R_1(\theta)$, $R_2(\theta)$, $R_3(\theta)$ and $R(\theta)$ are approximately  $[0, 0.6235]$, $[0, 0.7320]$, $[0, 0.5270]$ and $[0, 0.8217]$. Though the ranges seem to be quite wide, they can provide us useful information to be utilized next. 
\par 
To evaluate $s_i(\theta)$, the severity of each individual requirement violation, similarly we simulate 
 $
 \max_{e\in E}\max_{W\in U}\sum_{j=1}^k W_j  g_i(a^{(j)},e,\theta)\times I(g_i(a^{(j)},e,\theta) \geq 0).
 $
 The results for $s_1(\theta)$, $s_2(\theta)$ and $s_3(\theta)$ are respectively 0.1464, 0.0493 and 3.5989. Clearly the violation of $g_3$ is the most severe one while the violation of $g_2$ is the least.

\subsection{Rank for Uncertainties (C.3)}
\label{sec:c3}
Our analysis on the rank for epistemic uncertainties is based on the range of $R(\theta)$ obtained above. In our computation, we obtain $\min_{W\in U}/\max_{W\in U}\sum_{j=1}^k W_j I(g_i(a^{(j)},e,\theta)\geq 0\text{\ for some\ }i=1,2,3)$ for each eligible $e\in E$. For simplicity, we use $R_{min}$ and $R_{max}$ to denote these two values for each eligible $e\in E$ respectively. 

Our approach is to scrutinize the plots of $R_{min}$ and $R_{max}$ against each epistemic variable (not shown here due to space limit). For $R_{min}$, large value is notable, since it means that any distribution that provides similarity to the original data is going to fail with large probability. Therefore the most ideal reduction is to avoid the region of $e$ such that all $R_{min}$'s are large. For $R_{max}$, the largest $R_{max}$ for the region denotes the maximum failure probability that one can have. So we pay attention to the epistemic variables that could potentially reduce the ``worst-case'' failure probability. Based on these considerations, we conclude that the rank for epistemic uncertainties is $e_3>e_1>e_2>e_4$.

\section{Reliability-Based Design (Problem D)} \label{sec:D}
To find a reliability-optimal design point $\theta_{new}$, we minimize 
\begin{equation}
\max_{e\in E}\min_{W\in U}\sum_{j=1}^k W_jI(g(a^{(j)},e,\theta)\geq 0).\label{opt criterion}
\end{equation}
Here is the reason why we choose this function as the objective. For an eligible $e\in E$, if $\min_{W\in U}\sum_{j=1}^k W_jI(g(a^{(j)},e,\theta)\geq 0)$ is large, then the true probability in which the system fails must be even larger than this ``best-case" estimate, which implies that this point $e$ has a considerable failure likelihood.  The objective above thus aims to find a design point to minimize this best-case estimate, but taking the worst-case among all the eligible $e$'s. Arguably, one can use other criteria such as minimizing $\max_{e\in E}\max_{W\in U}\sum_{j=1}^k W_jI(g(a^{(j)},e,\theta)\geq 0)$, but this could make our procedure more conservative.
\par 
The optimization problem \eqref{opt criterion} is of a ``black-box" nature since the function $g$ is only observed through simulation, and the problem is easily non-convex. Our approach is to use a gradient descent to guide us towards a better $\theta_{new}$, with a goal of finding a reasonably good $\theta_{new}$ (instead of insisting on full optimality which could be difficult to achieve in this problem). Note that we need to sample $a^{(j)}$ when we land at a new $\theta$ during our iterations, and hence our approach takes the form of a stochastic gradient descent or stochastic approximation. Moreover, the gradient cannot be estimated in an unbiased fashion as we only have black-box function evaluation, and thus we need to resort to the use of finite-difference. This results in a zeroth-order or the so-called Kiefer-Wolfowitz (KW) algorithm. As we have a nine-dimensional design variable, we choose to update $\theta$ via a coordinate descent, namely at each iteration we choose one of the dimensions and run a central finite-difference along that dimension, followed by a movement of $\theta$ guided by this gradient estimate with a suitable step size. The updates are done in a round-about fashion over the dimensions. The perturbation size in the finite-difference is chosen of order $1/n^{1/4}$ here as it appears to perform well empirically (though theoretically other scaling could be better).
\par
\setlength{\textfloatsep}{0cm}
\begin{algorithm}[t]

    \caption{KW algorithm to find $\theta_{new}$}
    \textbf{Input: } The baseline design point $\theta_{baseline}$. The initial step size $c_0$. The initial perturbation size $a_0$. The max iteration $N_{max}$. The objective function $f(\theta)$.
    
    \textbf{Procedure:}
    \begin{algorithmic}
    
    \State Set $x_{now} = 1_9$ and $n=1$.
    
    \While{$n\leq N_{max}$}
    \State Set $c_n=c_0/n^{1/4}$ and $a_n=a_0/n$.
    
    \For{$i$ from 1 to 9}
    \State $u=f(\theta_{baseline}\circ(x_{now}+c_ne_i))$.
    
    \State $l=f(\theta_{baseline}\circ(x_{now}-c_ne_i))$.
    
    \State $g=(u-l)/(2c_n)$.
    
    \State $x_{now}=x_{now}-a_ng$.
    \EndFor
    \State $n=n+1$.
    \EndWhile
    \State Output $\theta_{baseline}\circ x_{now}$.
    \State ($\circ$ denotes the Hadamard product).
    \end{algorithmic}
    \label{opt}
\end{algorithm}
\setlength{\floatsep}{0cm}
Algorithm \ref{opt} shows the details of our optimization procedure. Considering that the components of $\theta_{baseline}$ are of very different magnitudes, we first perform a normalization to ease this difference. The quantity $x_{now}$ encodes the position of the normalized $\theta_{now}$, and $1_9$ denotes a  nine-dimensional vector of $1$'s that is set as the initial normalized design point. We set $c_0=a_0=0.1$ and $N_{max}=8$.
 \par 
After running the algorithm, we arrive at a new design point. Compared with the baseline design, the objective function decreases from 0.3656 to 0.2732. Note that this means that the best-case estimate of the failure probability, among the worst possible of all eligible $e$'s, is 0.2732.



For $\theta_{new}$, the ranges of $R_1(\theta)$, $R_2(\theta)$, $R_3(\theta)$ and $R(\theta)$ (defined in Section \ref{sec:c1c2}) are approximately  $[0, 0.5935]$, $[0, 0.7469]$, $[0, 0.5465]$ and $[0, 0.8205]$. 
We could observe from the plots of $R_{min}$ and $R_{max}$ that $e_2$ has significant different patterns on high values in both plots. According to the trends shown in the plots, we rank the epistemic variables as $e_2>e_3>e_1>e_4$.


\section{Design Tuning (Problem E)}
With data sequence $D_2=\{z^{(i)}(t)\}$ for $i=1,\dots,n_2$, we may incorporate the additional information to update our model as before, and we determine to refine $e_2$. The final design $\theta_{final}$ is obtained using Algorithm \ref{opt} with this updated information. 



\bibliographystyle{chicago} 
\bibliography{bibliography} 

\begin{thebibliography}{}

\bibitem[\protect\citeauthoryear{Asmussen and Glynn}{Asmussen and
  Glynn}{2007}]{asmussen2007stochastic}
Asmussen, S. and P.~W. Glynn (2007).
\newblock {\em Stochastic simulation: algorithms and analysis}, Volume~57.
\newblock Springer Science \& Business Media.

\bibitem[\protect\citeauthoryear{Ben-Tal, Den~Hertog, De~Waegenaere, Melenberg,
  and Rennen}{Ben-Tal et~al.}{2013}]{ben2013robust}
Ben-Tal, A., D.~Den~Hertog, A.~De~Waegenaere, B.~Melenberg, and G.~Rennen
  (2013).
\newblock Robust solutions of optimization problems affected by uncertain
  probabilities.
\newblock {\em Management Science\/}~{\em 59\/}(2), 341--357.

\bibitem[\protect\citeauthoryear{Ben-Tal and Nemirovski}{Ben-Tal and
  Nemirovski}{2002}]{ben2002robust}
Ben-Tal, A. and A.~Nemirovski (2002).
\newblock Robust optimization--methodology and applications.
\newblock {\em Mathematical Programming\/}~{\em 92\/}(3), 453--480.

\bibitem[\protect\citeauthoryear{Bertsimas, Brown, and Caramanis}{Bertsimas
  et~al.}{2011}]{bertsimas2011theory}
Bertsimas, D., D.~B. Brown, and C.~Caramanis (2011).
\newblock Theory and applications of robust optimization.
\newblock {\em SIAM Review\/}~{\em 53\/}(3), 464--501.

\bibitem[\protect\citeauthoryear{Bertsimas, Gupta, and Kallus}{Bertsimas
  et~al.}{2018a}]{bertsimas2018data}
Bertsimas, D., V.~Gupta, and N.~Kallus (2018a).
\newblock Data-driven robust optimization.
\newblock {\em Mathematical Programming\/}~{\em 167\/}(2), 235--292.

\bibitem[\protect\citeauthoryear{Bertsimas, Gupta, and Kallus}{Bertsimas
  et~al.}{2018b}]{bertsimas2018robust}
Bertsimas, D., V.~Gupta, and N.~Kallus (2018b).
\newblock Robust sample average approximation.
\newblock {\em Mathematical Programming\/}~{\em 171\/}(1-2), 217--282.

\bibitem[\protect\citeauthoryear{Blanchet and Lam}{Blanchet and
  Lam}{2012}]{blanchet2012state}
Blanchet, J. and H.~Lam (2012).
\newblock State-dependent importance sampling for rare-event simulation: An
  overview and recent advances.
\newblock {\em Surveys in Operations Research and Management Science\/}~{\em
  17\/}(1), 38--59.

\bibitem[\protect\citeauthoryear{Crespo and Kenny}{Crespo and
  Kenny}{2020}]{nasa}
Crespo, L. and S.~Kenny (2020).
\newblock The {NASA} {L}angley {C}hallenge on {O}ptimization under
  {U}ncertainty.
\newblock {\em ESREL\/}.

\bibitem[\protect\citeauthoryear{Delage and Ye}{Delage and
  Ye}{2010}]{delage2010distributionally}
Delage, E. and Y.~Ye (2010).
\newblock Distributionally robust optimization under moment uncertainty with
  application to data-driven problems.
\newblock {\em Operations research\/}~{\em 58\/}(3), 595--612.

\bibitem[\protect\citeauthoryear{Esfahani and Kuhn}{Esfahani and
  Kuhn}{2018}]{esfahani2018data}
Esfahani, P.~M. and D.~Kuhn (2018).
\newblock Data-driven distributionally robust optimization using the
  wasserstein metric: Performance guarantees and tractable reformulations.
\newblock {\em Mathematical Programming\/}~{\em 171\/}(1-2), 115--166.

\bibitem[\protect\citeauthoryear{Ghosh and Lam}{Ghosh and
  Lam}{2019}]{ghosh2019robust}
Ghosh, S. and H.~Lam (2019).
\newblock Robust analysis in stochastic simulation: Computation and performance
  guarantees.
\newblock {\em Operations Research\/}~{\em 67\/}(1), 232--249.

\bibitem[\protect\citeauthoryear{Glasserman and Xu}{Glasserman and
  Xu}{2014}]{glasserman2014robust}
Glasserman, P. and X.~Xu (2014).
\newblock Robust risk measurement and model risk.
\newblock {\em Quantitative Finance\/}~{\em 14\/}(1), 29--58.

\bibitem[\protect\citeauthoryear{Goeva, Lam, Qian, and Zhang}{Goeva
  et~al.}{2019}]{goeva2019optimization}
Goeva, A., H.~Lam, H.~Qian, and B.~Zhang (2019).
\newblock Optimization-based calibration of simulation input models.
\newblock {\em Operations Research\/}~{\em 67\/}(5), 1362--1382.

\bibitem[\protect\citeauthoryear{Hong, Huang, and Lam}{Hong
  et~al.}{2017}]{hong2017learning}
Hong, L.~J., Z.~Huang, and H.~Lam (2017).
\newblock Learning-based robust optimization: Procedures and statistical
  guarantees.
\newblock {\em arXiv preprint arXiv:1704.04342\/}.

\bibitem[\protect\citeauthoryear{Lam}{Lam}{2016}]{lam2013robust}
Lam, H. (2016).
\newblock Robust sensitivity analysis for stochastic systems.
\newblock {\em Mathematics of Operations Research\/}~{\em 41\/}(4), 1248--1275.

\bibitem[\protect\citeauthoryear{Wiesemann, Kuhn, and Sim}{Wiesemann
  et~al.}{2014}]{wiesemann2014distributionally}
Wiesemann, W., D.~Kuhn, and M.~Sim (2014).
\newblock Distributionally robust convex optimization.
\newblock {\em Operations Research\/}~{\em 62\/}(6), 1358--1376.

\end{thebibliography}
\newpage
\end{document}